\documentclass[a4paper,conference]{IEEEtran}      

\IEEEoverridecommandlockouts                              

\usepackage{graphics} 
\usepackage{epsfig} 
\usepackage{times} 
\usepackage{amsmath} 
\usepackage{amssymb}  
\usepackage{amsfonts}
\usepackage{subfigure}
\usepackage{url}
\usepackage{bbm, dsfont}
\usepackage{mathtools}
\newtheorem{theorem}{Theorem}

\newtheorem{algorithm}[theorem]{Algorithm}

\usepackage{cite}

\usepackage{color}
\usepackage{algorithm}

\newtheorem{definition}[theorem]{Definition}

{ 

}

\newcommand{\bi}{\begin{itemize}}
\newcommand{\ei}{\end{itemize}}
\newcommand{\bd}{\begin{displaymath}}
\newcommand{\ed}{\end{displaymath}}
\newcommand{\be}{\begin{eqnarray*}}
\newcommand{\ee}{\end{eqnarray*}}

\usepackage{textcomp}
\usepackage{xcolor}
\def\BibTeX{{\rm B\kern-.05em{\sc i\kern-.025em b}\kern-.08em
    T\kern-.1667em\lower.7ex\hbox{E}\kern-.125emX}}

\begin{document}

\title{Computationally Efficient Learning of Large Scale Dynamical Systems: A Koopman Theoretic Approach*\\
\thanks{This work was supported partially by a Defense Advanced Research Projects Agency (DARPA) Grant No. DEAC0576RL01830 and Institute of Collaborative Biotechnologies Grant. }
}

\author{\IEEEauthorblockN{1\textsuperscript{st} Sinha, Subhrajit}
\IEEEauthorblockA{\textit{Biological Science Facility} \\
\textit{Pacific Northwest National Laboratory}\\
Richland, USA \\
subhrajit.sinha@pnnl.gov}
\and
\IEEEauthorblockN{2\textsuperscript{nd} Nandanoori, Sai Pushpak}
\IEEEauthorblockA{\textit{Control and Optimization Group} \\
\textit{Pacific Northwest National Laboratory}\\
Richland, USA \\
saipushpak.n@pnnl.gov}
\and
\IEEEauthorblockN{3\textsuperscript{rd} Yeung, Enoch}
\IEEEauthorblockA{\textit{Mechanical Engineering Department} \\
\textit{University of California, Santa Barbara}\\
Santa Barbara, USA\\
eyeung@ucsb.edu}
}

\maketitle

%
%
%
%
%
%

\begin{abstract}
In recent years there has been a considerable drive towards data-driven analysis, discovery and control of dynamical systems. To this end, operator theoretic methods, namely, Koopman operator methods have gained a lot of interest. In general, the Koopman operator is obtained as a solution to a least-squares problem, and as such, the Koopman operator can be expressed as a closed-form solution that involves the computation of Moore-Penrose inverse of a matrix. For high dimensional systems and also if the size of the obtained data-set is large, the computation of the Moore-Penrose inverse becomes computationally challenging. In this paper, we provide an algorithm for computing the Koopman operator for high dimensional systems in a time-efficient manner. We further demonstrate the efficacy of the proposed approach on two different systems, namely a network of coupled oscillators (with state-space dimension up to 2500) and IEEE 68 bus system (with state-space dimension 204 and up to 24,000 time-points).
\end{abstract}

\section{Introduction}\label{section_Introduction}
Dynamical systems theory dates back to Newton when he came up with the laws of motion. Since then, it has flourished as a branch of mathematics and physics, with applications to a plethora of scientific and engineering disciplines. In recent times, with advancements in computational and data handling capacity, data-driven analysis of dynamical systems has gained a real impetus. 

In the realm of data-driven analysis of dynamical systems, one of the most popular methods is based on operator theoretic methods, namely, the Koopman and Perron-Frobenius (P-F) operators \cite{Lasota}. Given a dynamical system, P-F and Koopman operators are typically infinite-dimensional operators which govern the evolution of distributions (or measures) and functions, respectively,  under the system dynamics \cite{Lasota,Vaidya_TAC,Mezic_comparison,mezic2005spectral,Mehta_comparsion_cdc,mezic_koopmanism}. Though these operators are infinite-dimensional, one major advantage is the fact that these are linear operators and hence even if the underlying system is nonlinear, one obtains a linear system, albeit infinite-dimensional \cite{Lasota}. It is the linearity property that has driven researchers to explore these operator theoretic ideas for data-driven analysis of dynamical systems.

Data-driven discovery and control of dynamical systems, using the P-F and Koopman operators involves the construction of finite-dimensional approximations of these operators. Computation of P-F operators involves set-theoretic methods \cite{Junge_Osinga,sinha_optimal_placement_ecc,sinha_optimal_placement} and is typically computationally expensive. Finite-dimensional approximation of Koopman operator, on the other hand, is computed by studying the evolution of \emph{observables} and usually involves finding a solution to a least-squares problem and hence is computationally much less expensive. Moreover, a Koopman operator can be trained on only a few trajectories and thus, one does not require a large number of initial conditions. Hence, the Koopman operator is more popular for data-driven analysis of dynamical systems. This has resulted in a huge drive in the development of Koopman operator methods for dynamical systems analysis and control \cite{Dellnitz_Junge,Mezic2000,froyland_extracting,Junge_Osinga,Mezic_comparison,
Dellnitztransport,mezic2005spectral,Mehta_comparsion_cdc,Vaidya_TAC,EDMD_williams,susuki2011nonlinear,mezic_koopmanism,yeung2018koopman,yeung2017learning,sinha_robust_DMD_acc,sinha_robust_DMD_journal,sparse_Koopman_acc,sinha_equivariant,sinha_online_koopman_arxiv,sinha_online_PES,sai_global_koopman_acc}.

Among the different methods for computation of the finite-dimensional approximation of the Koopman operator, the most popular methods are Dynamic Mode Decomposition (DMD) \cite{mezic2005spectral} and Extended Dynamic Mode Decomposition (EDMD) \cite{EDMD_williams}. Typically the trajectories are lifted to a higher dimensional feature space and the finite-dimensional Koopman operator is obtained as a solution of a norm minimization problem on the feature space and as such when the norm considered is the 2-norm, the Koopman operator can be computed directly using the Moore-Penrose inverse. However, for high dimensional systems computation of the Moore-Penrose and hence the Koopman operator becomes computationally challenging. In particular, the standard algorithms for computation of the Moore-Penrose inverse involves Singular Value Decomposition (SVD) and the computation is of the order $O(n^3)$. Moreover, if the number of data-points is large, it poses similar challenges.

In this paper, we provide a new method for computing the Koopman operator for high dimensional systems (and/or large data sets) in a time-efficient manner. In particular, the method is based on Cholesky decomposition \cite{courrieu2008fast} so that the dimension of the matrix being inverted is reduced. Furthermore, if parallel processing capabilities is utilised, the computation time is reduced further, thus making this algorithm for data-driven identification of dynamical systems using Koopman operator easily scalable. To demonstrate the efficacy of the proposed approach, we consider two different dynamical systems wherein one example we increase the dimension of the system (keeping the number of data points constant) and compare the computation times, while in the other example we consider a high dimensional system and consider different sizes of the data-set.

The paper is organized as follows. In section \ref{section_Preliminaries}, we briefly review the concepts of P-F and Koopman operators, followed by the EDMD algorithm for the finite-dimensional approximation of the Koopman operator in section \ref{section_DMD}. In section \ref{section_Efficient}, we describe the algorithm for the efficient computation of the Koopman operator. In section \ref{section_Simulations}, we demonstrate the efficacy of the method by computing the Koopman operators for two high-dimensional systems, namely, a network of linear oscillators and IEEE 68 bus system. The paper is finally concluded in section \ref{section_Conclusions}.

\section{Preliminaries}\label{section_Preliminaries}
Consider a discrete-time dynamical system
\begin{align}\label{system}
x_{t+1} = T(x_t)
\end{align}
where $T: X\subseteq \mathbb{R}^N \to X$ is assumed to be at least ${\cal C}^1$.  Associated with the dynamical system (\ref{system}) is the Borel-$\sigma$ algebra ${\cal B}(X)$ on $X$ and the vector space ${\cal M}(X)$ of bounded complex valued measures on $X$. With this, two linear operators, namely, Perron-Frobenius (P-F) and Koopman operator, can be defined as follows \cite{Lasota}:
\begin{definition}
The P-F operator $\mathbb{P}:{\cal M}(X)\to {\cal M}(X)$ is given by
\[[\mathbb{P}\mu](A)=\int_{{X} }\delta_{T(x)}(A)d\mu(x)=\mu(T^{-1}(A))\]
$\delta_{T(x)}(A)$ is stochastic transition function which measure the probability that point $x$ will reach the set $A$ in one time step under the system mapping $T$. 
\end{definition}

%
%

\begin{definition} 
Given any $h\in\cal{F}(X)$, the Koopman operator $\mathbb{U}:{\cal F}(X)\to {\cal F}(X)$ is defined as
$[\mathbb{U} h](x)=h(T(x))$, where ${\cal F}(X)$ is the space of functions (observables), defined on $X$, invariant under the action of the Koopman operator.
\end{definition}

 \begin{figure}[htp!]
 \centering
 \includegraphics[scale=.23]{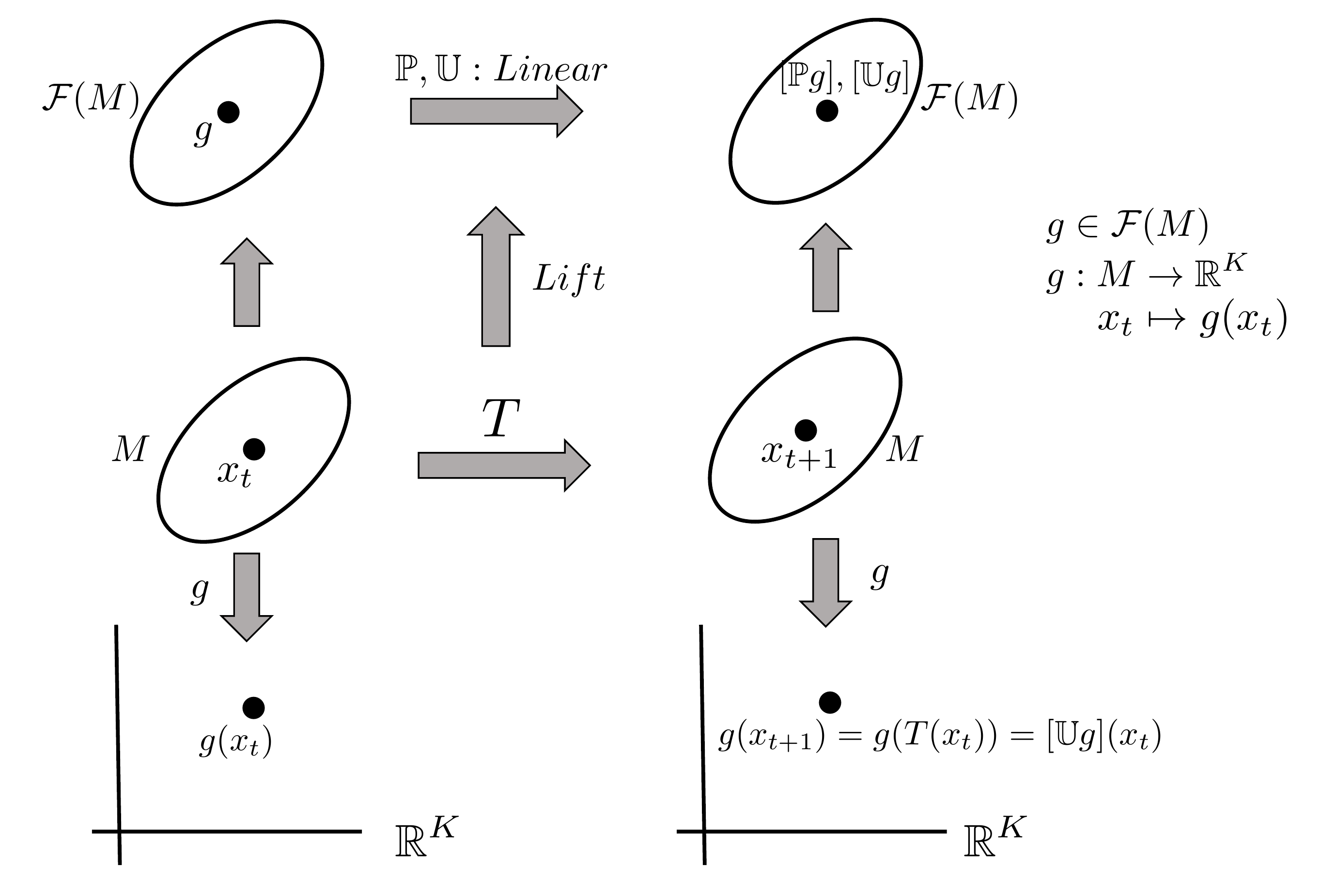}
 \caption{Schematic of the P-F and Koopman operators.}\label{koopman_diagram}
 \end{figure}

Both the Perron-Frobenius and the Koopman operators are linear operators, even if the underlying system is nonlinear. But while analysis is made tractable by linearity, the trade-off is that these operators are typically infinite-dimensional. In particular, the P-F operator and Koopman operator often will lift a dynamical system from a finite-dimensional space to generate an infinite-dimensional linear system.


\section{Data-Driven Discovery of Dynamical Systems}\label{section_DMD}
The P-F and Koopman operators are adjoint to each other, however, since the Koopman operator governs the evolution of functions under the system dynamics (instead of densities as in the case of P-F operators) it is more suitable for data-driven analysis and discovery of dynamical systems.

\subsection{Extended Dynamic Mode Decomposition}
In this subsection, we briefly describe the popular Extended Dynamic Mode Decomposition (EDMD) algorithm for the computation of the Koopman operator from time-series data \cite{EDMD_williams}. 

Let 
\begin{eqnarray}
X_p = [x_1,x_2,\ldots,x_M],& X_f = [y_1,y_2,\ldots,y_M] \label{data}
\end{eqnarray}
where $x_i\in X$ and $y_i\in X$, be snapshots of data obtained from a simulation of a system $x\mapsto T(x)$ or from an experiment . The two pair of data sets are assumed to be two consecutive snapshots i.e., $y_i=T(x_i)$. Let $\mathcal{D}=
\{\psi_1,\psi_2,\ldots,\psi_K\}$ be the set of dictionary functions or observables, where $\psi : X \to \mathbb{C}$. Let ${\cal G}_{\cal D}$ denote the span of ${\cal D}$ such that ${\cal G}_{\cal D}\subset {\cal G}$, where ${\cal G} = L_2(X,{\cal B},\mu)$. Define vector valued function $\mathbf{\Psi}:X\to \mathbb{C}^{K}$
\begin{equation}
\mathbf{\Psi}(\boldsymbol{x}):=\begin{bmatrix}\psi_1(x) & \psi_2(x) & \cdots & \psi_K(x)\end{bmatrix}^\top
\end{equation}
Any function $\phi,\hat{\phi}\in \mathcal{G}_{\cal D}$ can be written as
\begin{eqnarray}
\phi = \sum_{k=1}^K a_k\psi_k=\boldsymbol{\Psi}^\top a,\quad \hat{\phi} = \sum_{k=1}^K \hat{a}_k\psi_k=\boldsymbol{\Psi}^\top \hat{a}
\end{eqnarray}
for some set of coefficients $\boldsymbol{a},\boldsymbol{\hat{a}}\in \mathbb{C}^K$. Let \[ \hat{\phi}(x)=[\mathbb{U}\phi](x)+r,\]
where $r$ is a residual function that appears because $\mathcal{G}_{\cal D}$ is not necessarily invariant to the action of the Koopman operator. Let $\bf K$ be the finite dimensional approximation of the Koopman operator. Then the  matrix $\bf K$ is obtained as a solution of least square problem as follows 
\begin{equation}\label{edmd_op}
\min\limits_{\bf K}\parallel {\bf K}Y_p  - Y_f\parallel_2
\end{equation}
where
\begin{eqnarray}\label{edmd1}
\begin{aligned}
&{Y_p}={\bf \Psi}(X_p) = & [{\bf \Psi}(x_1), {\bf \Psi}(x_2), \cdots , {\bf \Psi}(x_M)]\\
& {Y_f}={\bf \Psi}(X_f) = & [{\bf \Psi}(y_1), {\bf \Psi}(y_2), \cdots , {\bf \Psi}(y_M)],
\end{aligned}
\end{eqnarray}
 with ${\bf K}\in\mathbb{C}^{K\times K}$. The optimization problem (\ref{edmd_op}) can be solved using a multitude of techniques like batch gradient descent, stochastic gradient descent or can be solved explicitly to obtain the following solution for the matrix $\bf K$
\begin{eqnarray}
{\bf K}_{EDMD}= Y_fY_p^\dagger\label{EDMD_formula}
\end{eqnarray}
where ${\bf G}^{\dagger}$ is the Moore-Penrose of matrix $\bf G$.
DMD is a special case of EDMD algorithm with ${\bf \Psi}(x) = x$.

\section{Efficient Computation of Koopman Operator}\label{section_Efficient}

Since the Koopman operator is obtained as a solution of a least-squares problem, one can solve it directly by computing the Moore-Penrose inverse of a suitable matrix. The Moore-Penrose inverse of a $p\times q$ matrix $M$ is a unique $q\times p$ matrix ${M}^\dagger$, such that
\[M{M}^\dagger M=M;\quad{M}^\dagger M{M}^\dagger={M}^\dagger.\]
When $M$ is of full rank the Moore-Penrose inverse is the usual pseudo-inverse $M^\dagger= (M^\top M)^{-1}M^\top$. However, when $M$ is rank deficient, the computation of $M^\dagger$ is not straightforward and there are several algorithms to compute the Moore-Penrose inverse \cite{ben2003generalized}, with the Singular Value Decomposition (SVD) technique being one of the most commonly used algorithms. However, the computational time of these algorithms grows rapidly with the size of the matrix $M$. 

With this said, the EDMD algorithm usually suffers from the curse of dimensionality. The key feature of the EDMD algorithm is to choose a set of dictionary functions $\cal D$, such that the dictionary functions are rich enough to approximate the Koopman invariant subspaces. To achieve this, usually, the cardinality of the set $\cal D$ is large and hence the dimension of the feature space is large, even if the dimension of the underlying system is small. As such, the number of dictionary functions increases rapidly with the dimension of the underlying system. Hence, the computation of the Koopman operator becomes slow and is thus computationally inefficient. Furthermore, often the matrices $Y_p$ and $Y_f$, as defined in (\ref{edmd1}), are rank deficient, especially if most of the data points $x_i$ are close to an attractor set. Hence, computing the Koopman operator using the existing algorithms becomes quite complex.

\subsection{The algorithm}

Let $M^\top M$ be a $q\times q$ symmetric matrix so that its rank is $r\leq q$. Hence, there exists a unique upper triangular matrix $U$ with $(q-r)$ zero rows such that
\[U^\top U = M^\top M.\]
Hence, removal of the zero rows results in a matrix $L\in\mathbb{R}^{r\times q}$ with full rank $r$. Moreover,
\[U^\top U = M^\top M=LL^\top.\]
With this we have
\begin{theorem}
\begin{equation}
M^\dagger = L (L^\top L)^{-1}(L^\top L)^{-1}L^\top M^\top.
\end{equation}
\end{theorem}

\textbf{Proof.} From \cite{rakha2004moore}, we have for a product matrix $XY$, 
\[(XY)^\dagger = Y^\top(X^\top X Y Y^\top)^\dagger X^\top.\]
Hence, when $Y$ is the identity matrix, we have
\[X^\dagger = (X^\top X)^\dagger X^\top.\]
Moreover, if $B = X^\top$ and $X\in\mathbb{R}^{q\times r}$ is of rank $r$, then
\[(XX^\top)^\dagger = X(X^\top X)^{-1} (X^\top X)^{-1} X^\top.\]
Hence, 
\begin{eqnarray}\label{lower_inverse}
(M^\top M)^\dagger = (LL^\top)^\dagger = L(L^\top L)^{-1} (L^\top L)^{-1} L^\top.
\end{eqnarray}
Using, $X^\dagger = (X^\top X)^\dagger X^\top$, we have
\begin{eqnarray}\label{penrose_inverse}
\begin{aligned}
M^\dagger &= (M^\top M)^\dagger M^\top\\
& = L(L^\top L)^{-1} (L^\top L)^{-1} L^\top M^\top.
\end{aligned}
\end{eqnarray}
\hfill $\blacksquare$

Using the above theorem, the computation of Koopman operator can be summarized as follows:
\begin{algorithm}[h!]
\caption{Efficient Computation of Koopman Operator}
\begin{enumerate}
\item{From the obtained data set form the sets $X_p$ and $X_f$ as given in (\ref{data}).}
\item{Choose the dictionary functions ${\bf \Psi}(x)$.}
\item{Compute $Y_p$ and $Y_f$ as given by (\ref{edmd1}).}
\item{Using the Cholesky decomposition of $Y_p$, compute the Moore-Penrose inverse $(Y_p^\dagger)$ of $Y_p$ by the formula (\ref{penrose_inverse}).}
\item{Compute the Koopman Operator as ${\bf K}=Y_fY_p^\dagger$.}
\end{enumerate}
\label{algo}
\end{algorithm}

\subsection{Computational cost}

The above theorem provides an algorithm of finding the Moore-Penrose inverse of a rank deficient non-square matrix. Computationally, there are two steps that are computationally expensive, namely, computing full rank Cholesky decomposition of $M^\top M$ and computing the inverse of $L^\top L$. On a serial processor, these computations are of order $O(q^3)$ and $O(r^3)$, respectively. However, if one utilizes parallel processing, then the complexity is of order $O(q)$ \cite{voevodin2002voevodin}. Furthermore, even when using serial processing, inverting a $r\times r$ matrix is computationally less expensive than inverting a $q\times q$ matrix, where $r<q$. Thus, even in serial processing, the above algorithm is efficient and when parallel processing is available, the complexity of the inversion step is of order $O(\log r)$ \cite{courrieu2008solving}.

\section{Simulation Results}\label{section_Simulations}
In this section, we demonstrate the efficacy of the proposed approach over SVD based method for computation of the Koopman operator for a high-dimensional system. In one of the examples, namely a network of coupled oscillators, we vary the number of oscillators in the network, while keeping the number of time-points of the data constant, while in the power network example, we vary the number of data points, while keeping the number of states of the system constant. The rationale behind this approach is the fact that while computing the Koopman operator, the size of the matrix whose inverse is to be computed can become large because of one or both of two reasons, namely, the number of states is large or the number of data points is large. We address both the cases in the following simulations. All the simulations were performed in MATLAB R2019a on an Apple Macbook Pro with 2.3 GHz Intel Core i5 processor and 8 GB 2133 MHz LPDDR3 RAM.

\subsection{Network of Coupled Oscillators}

\begin{figure}[h!]
\centering
{\includegraphics[scale=.3]{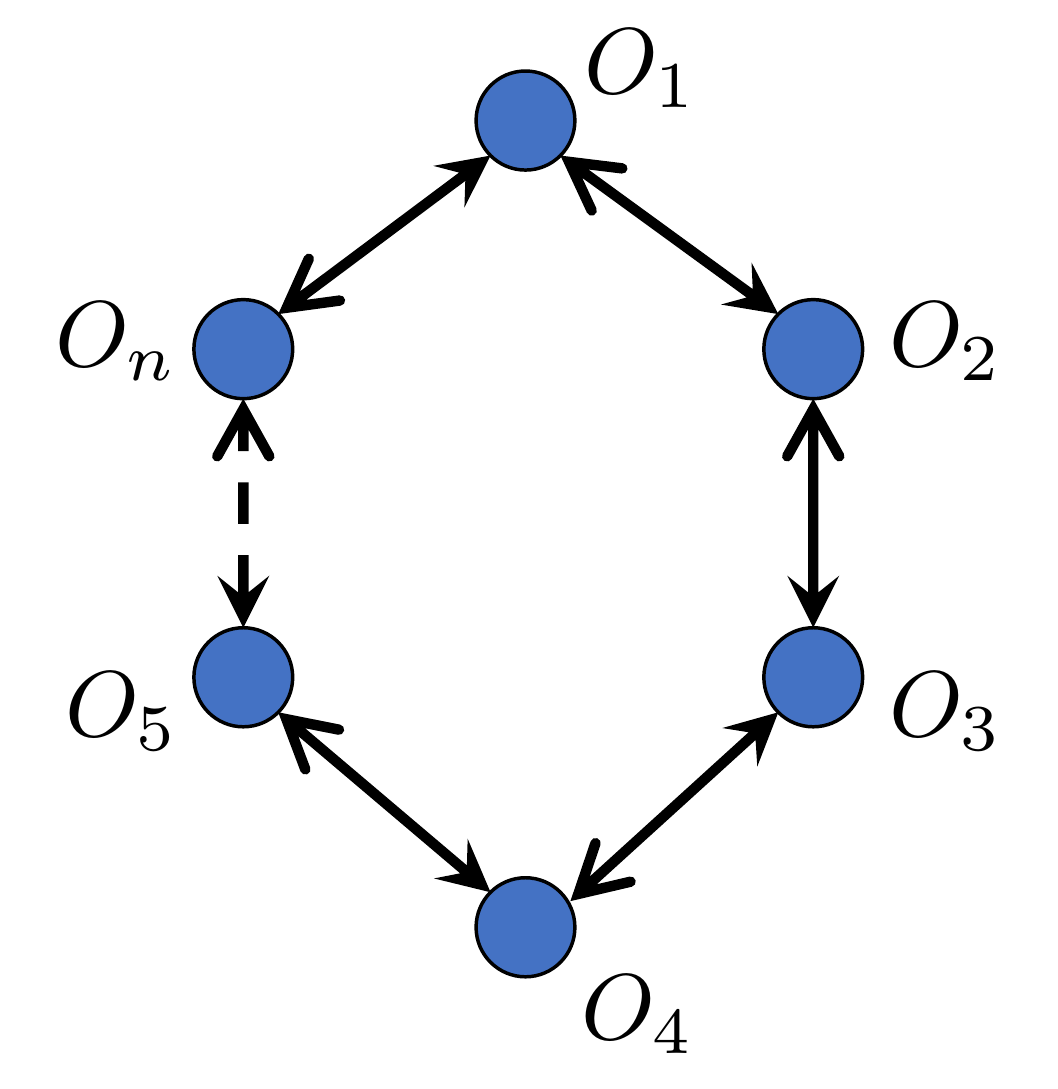}}
\caption{Ring network of second order oscillators.}\label{osc_network}
\end{figure}

Consider a network of coupled linear oscillators given by
\begin{eqnarray}\label{coup_osc}
\ddot{\theta}_k &=& -\mathcal{L}_k\theta - d\dot{\theta}_k, \quad k = 1,\cdots , N
\end{eqnarray}
where $\theta_k$ is the angular position of the $k^{th}$ oscillator, $N$ is the number of oscillators, $\mathcal{L}_k$ is the $k^{th}$ row of the Laplacian $\mathcal{L}$ and $d$ is the damping coefficient. In the simulations, the damping coefficient has been set to $0.4$ for all the oscillators and the data was collected with a sampling time of $\delta t = 0.01$ seconds. Furthermore, we assume a ring topology of the network as shown in Fig. \ref{osc_network}.

This network of oscillators has some similarities with the power network in the sense that each oscillator model follows closely with the second-order model of a synchronous generator and as they are connected over a network, it represents the interactions of the generators. The weighted Laplacian matrix shown in (\ref{coup_osc}) captures the admittance matrix of the power network. 
\begin{figure}[h!]
\centering
\subfigure[]{\includegraphics[scale=.228]{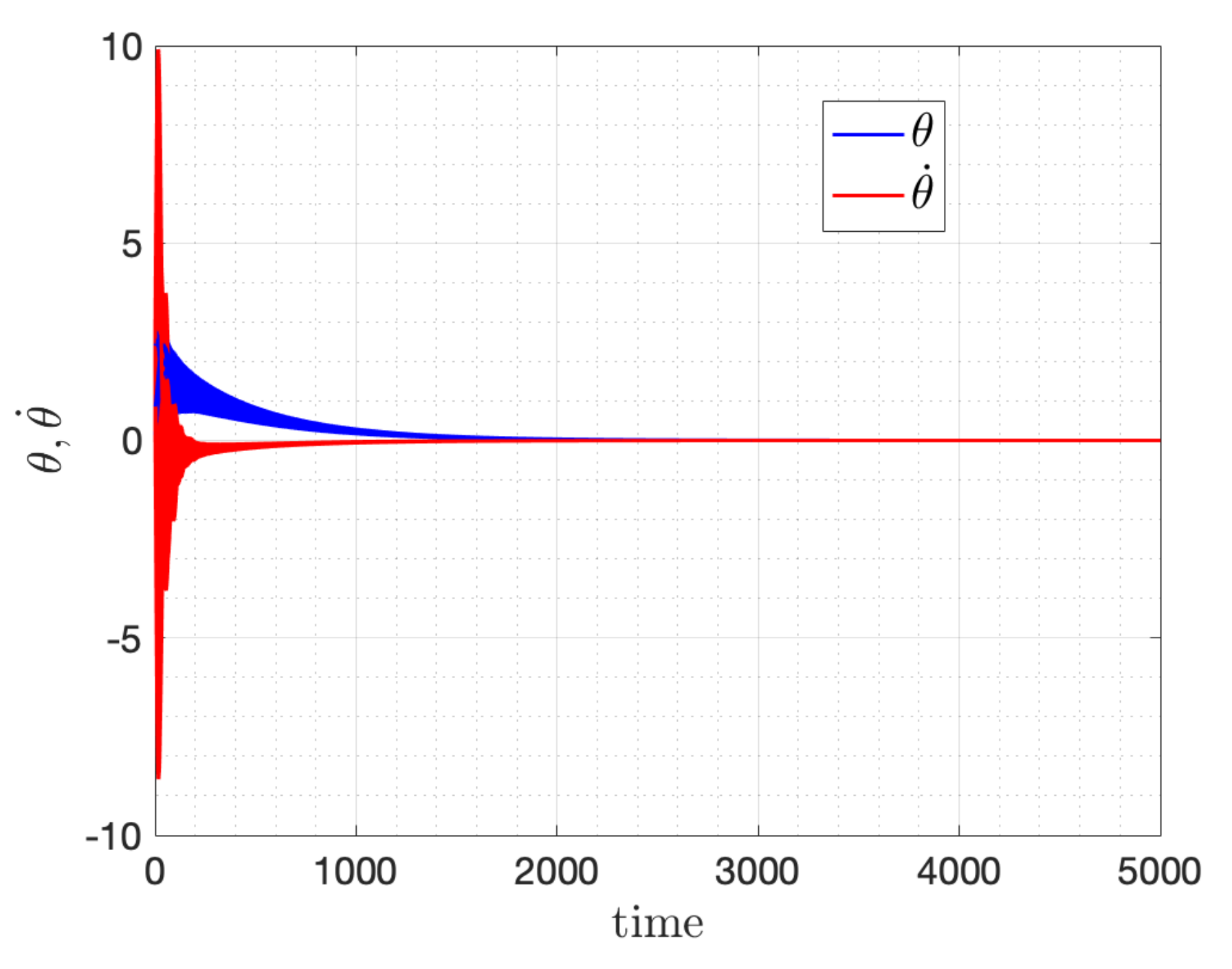}}
\subfigure[]{\includegraphics[scale=.252]{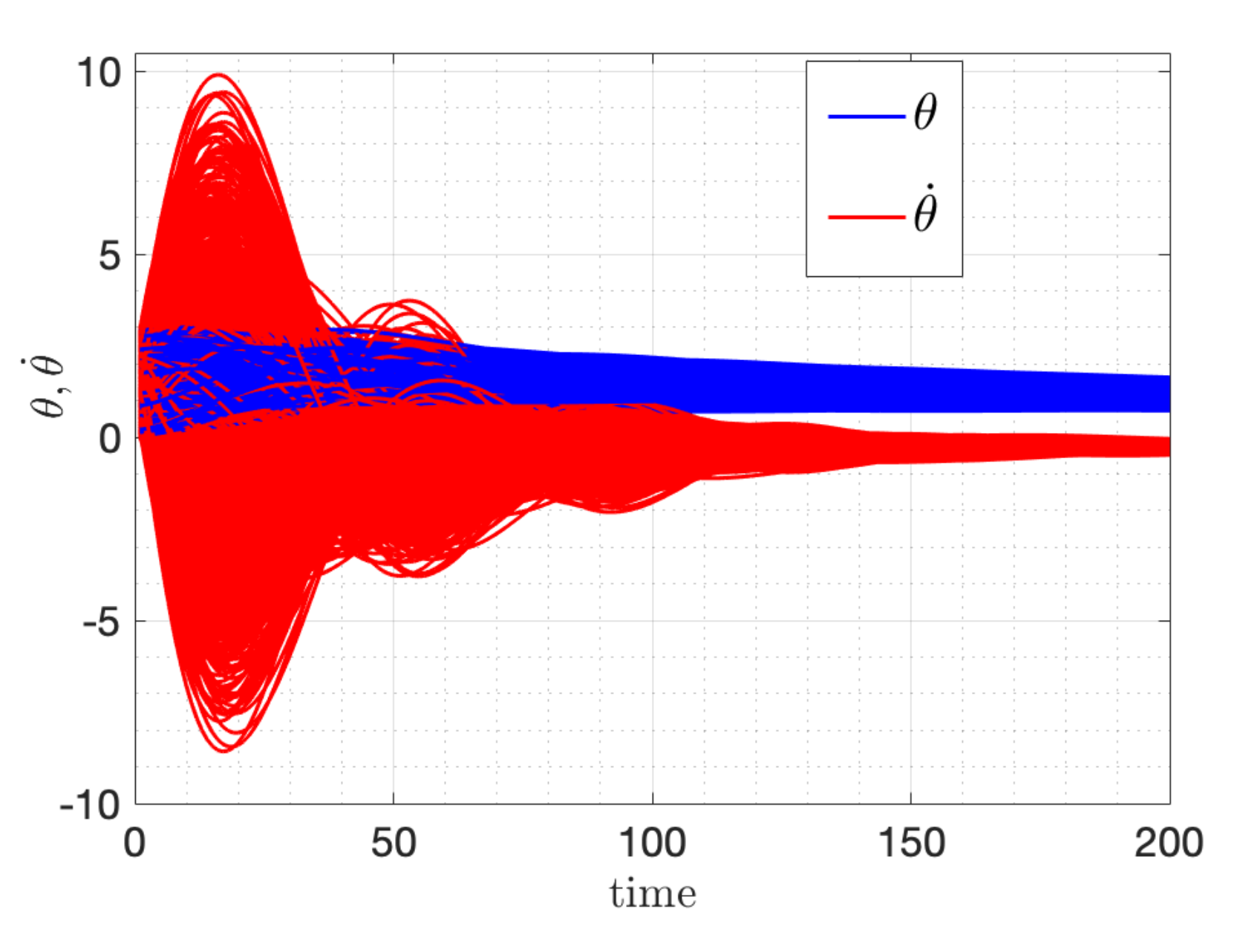}}
\caption{(a) Time domain trajectories of the oscillator network with 1250 oscillators. (b) Transient trajectories of the oscillator network with 1250 oscillators.}\label{osc_traj}
\end{figure}

\begin{figure}[h!]
\centering
\includegraphics[scale=.2]{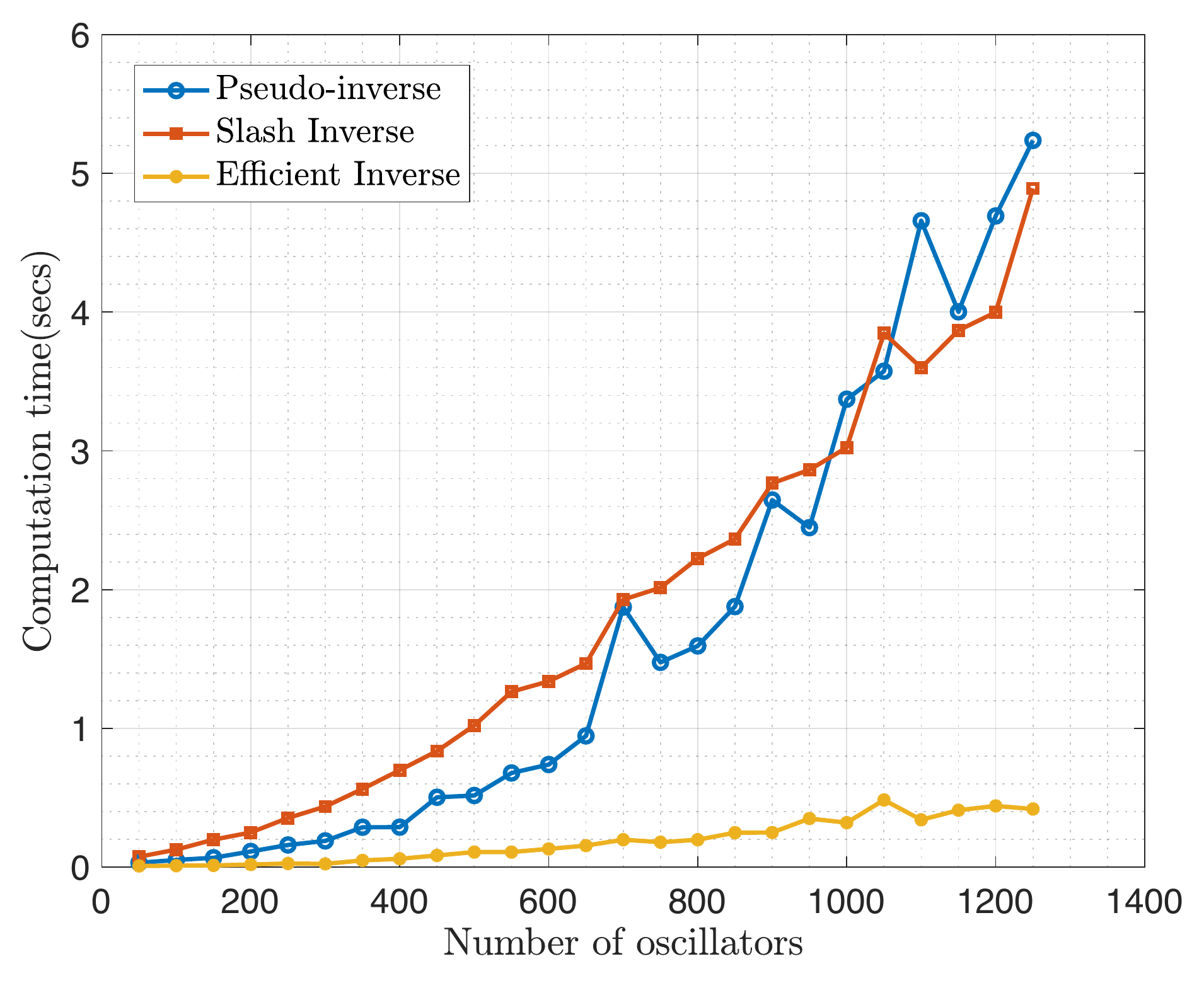}
\caption{Comparison of computation times of the Koopman operator for the oscillator network using different algorithms.}\label{osc_comp_time}
\end{figure} 

In the simulations, we vary the number of oscillators from 50 oscillators to 1250 oscillators, so that the dimension of the system varies from 100 to 2500. Furthermore, in all the simulations, data was collected for 5000 time steps. The time-domain trajectories of 1250 oscillators are shown in Fig. \ref{osc_traj}. Again, since the dynamical system is linear, we use linear dictionary functions, that is, ${\bf \Psi}(x)=x$ and as such the dimension of the feature space is equal to the dimension of the underlying system.

With the 5000 data points, we compute the Koopman operator for the various networks and compare the computation times of the various algorithms. In particular, we compare the efficiency of our algorithm with the existing MATLAB commands of {\tt pinv}, which uses SVD decomposition, and {``$\backslash$" (\tt backslash)}, which uses QR decomposition with pivoting.

The computation times of the different algorithms for computing the Koopman operator is shown in Fig. \ref{osc_comp_time}. As can be seen from the figure, the computation time of the Koopman operator is greatly reduced by the proposed method. Moreover, the trend is almost linear, whereas the other algorithms scale in a nonlinear fashion with the size of the underlying system dimension. As such, the proposed method will scale much more efficiently with the dimension of the system.

\begin{figure}[h!]
\centering
\subfigure[]{\includegraphics[scale=.2]{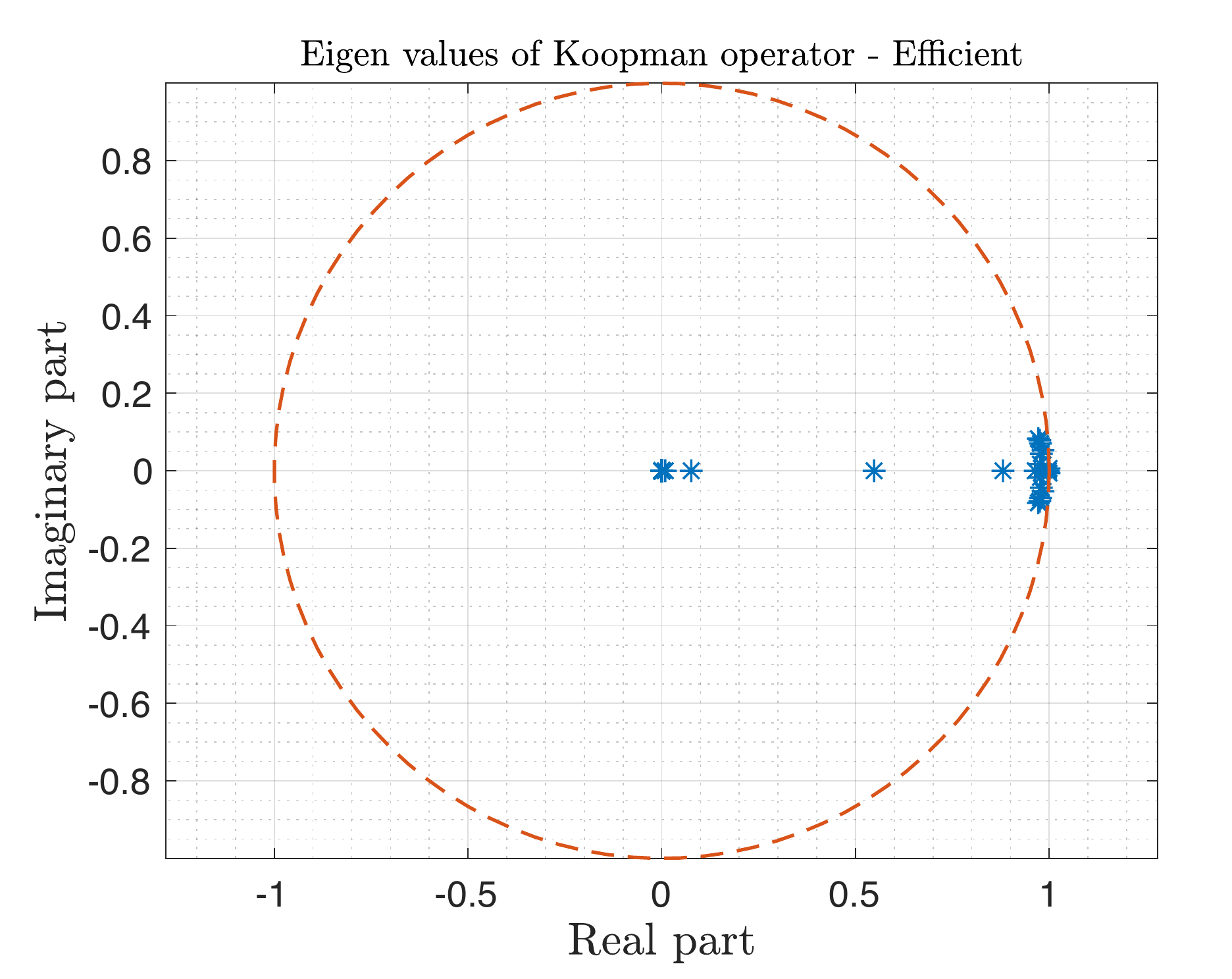}}
\subfigure[]{\includegraphics[scale=.2]{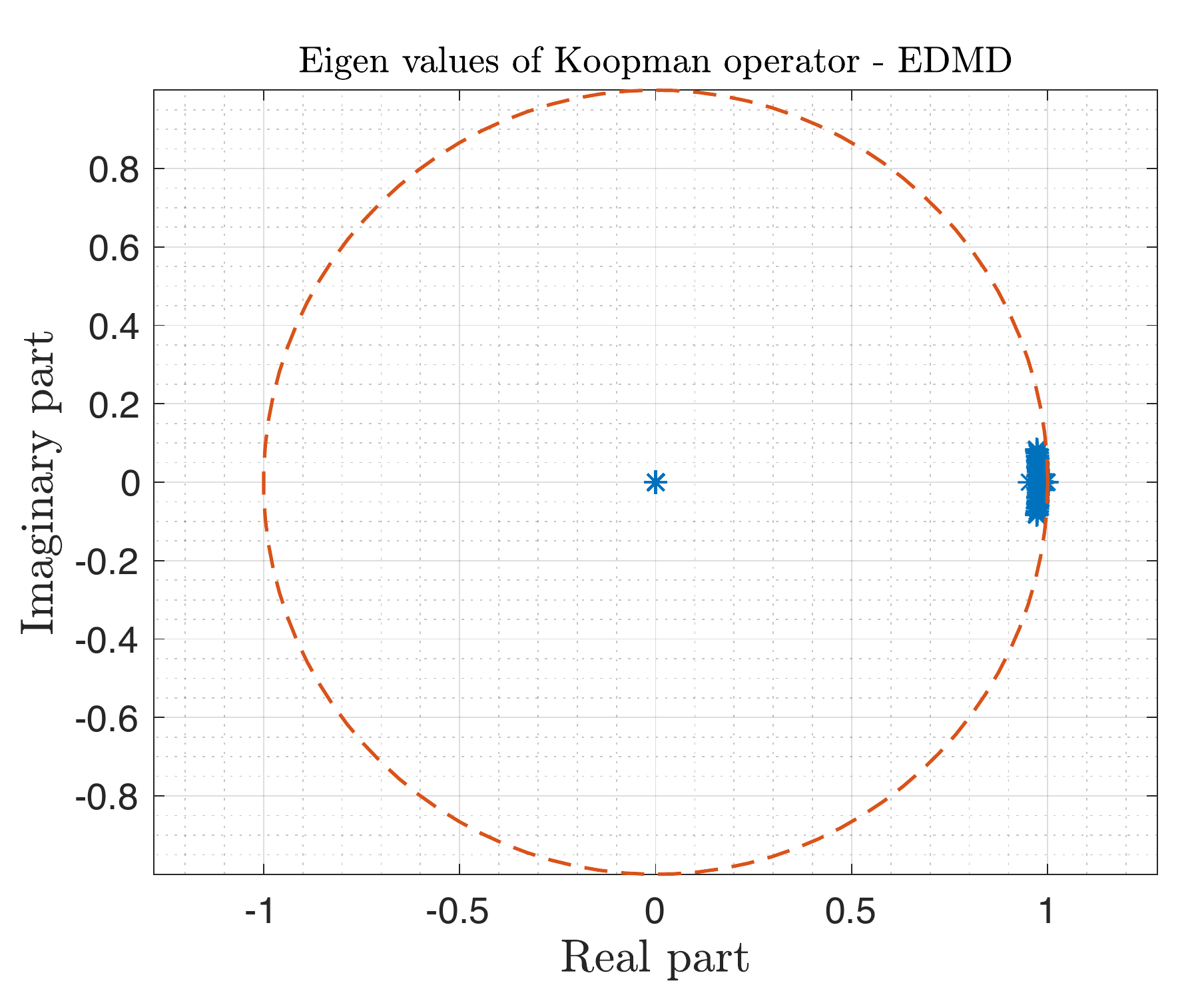}}
\caption{(a) Eigenvalues of the Koopman operator when computed using proposed approach. (b) Eigenvalues of the Koopman operator computed using the SVD based inverse computation ({\tt pinv} in MATLAB).}\label{osc_eig}
\end{figure}

The eigenvalues of the obtained Koopman operators (proposed method and {\tt pinv} method) for 750 oscillators are shown in Fig. \ref{osc_eig} and it can be seen that the dominant eigenvalues are similar using the two different techniques. As such, the long term behaviour, which is governed by the dominant eigenvalues, will be similar for both the Koopman operators. Note that the eigenvalue plots act as a validation of the obtained Koopman model.

\subsection{Discovery of Power Network}
\begin{figure}[h!]
\centering
\includegraphics[scale=.45]{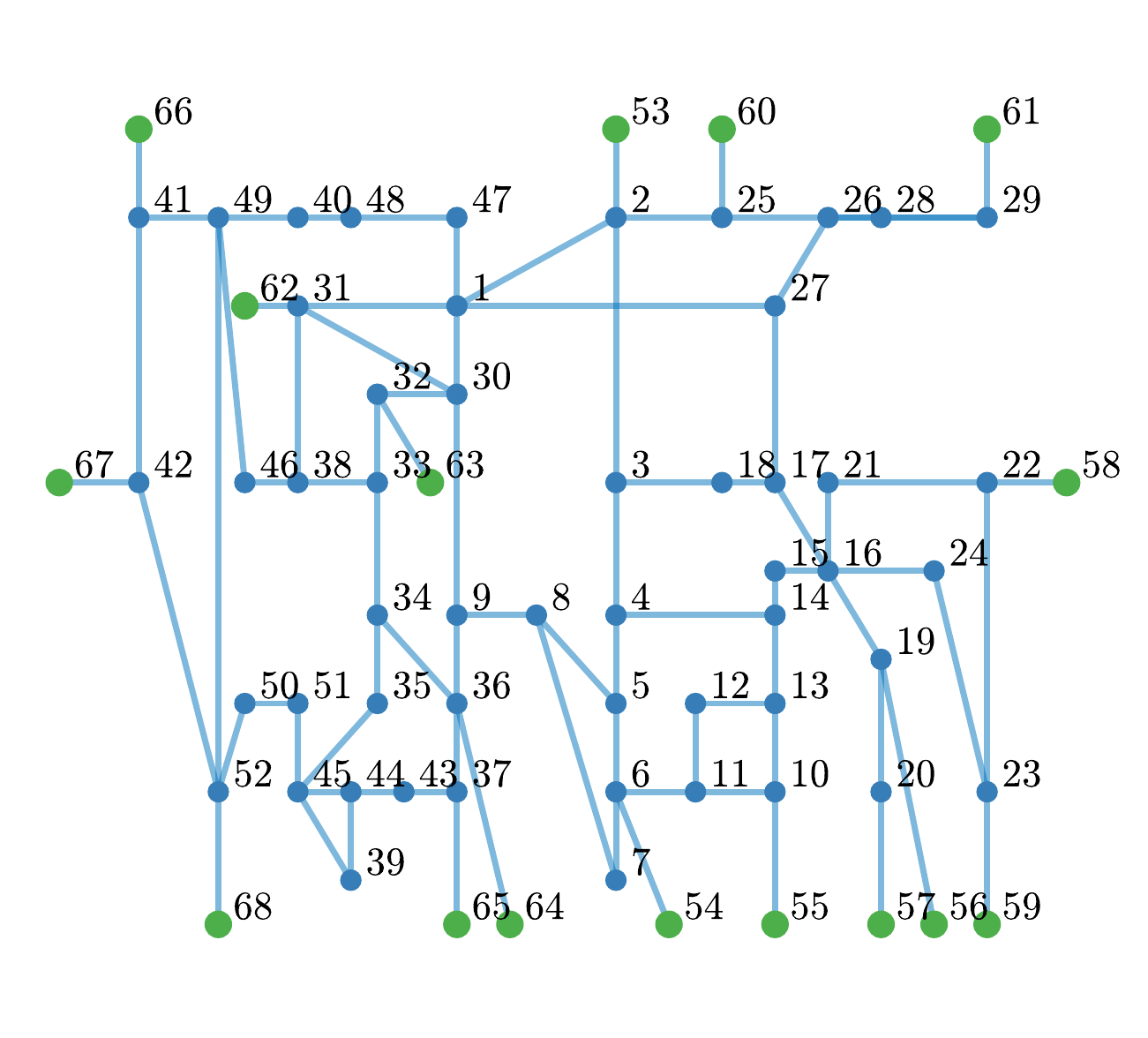}
\caption{IEEE 68 bus network topology. The green nodes correspond to the generators.}\label{power_network_topology}
\end{figure}

In this subsection, we demonstrate the proposed algorithm on the real-time PMU data. \textit{GridSTAGE} \cite{gridstage2020} simulation platform developed by Pacific Northwest National Laboratory (PNNL) under the support of Department of Energy's Advanced Grid Modeling program is leveraged to generate the synthetic data for a power network. GridSTAGE is developed based on Power System Toolbox (PST) \cite{sauer2017power} where nonlinear time-domain simulations can be generated for standard IEEE bus systems. GridSTAGE emulates PMU, SCADA sensors and provides rich multivariate spatio-temporal network data. In particular, a IEEE $68$ bus system is considered (Fig. \ref{power_network_topology}) and random load changes were considered at random bus locations with small magnitudes to generate transient data for the power network. \textit{PMU} data whose resolution is $50$ measurements/second is generated and for the sake of convenience, we assume a PMU is available at each bus. PMU data contains nodal measurements such as voltage magnitude (in p.u.), voltage angle, frequency and rate of change of frequency along with branch measurements such as current magnitude and current angle. 

\begin{figure}[h!]
\centering
\includegraphics[scale=.4]{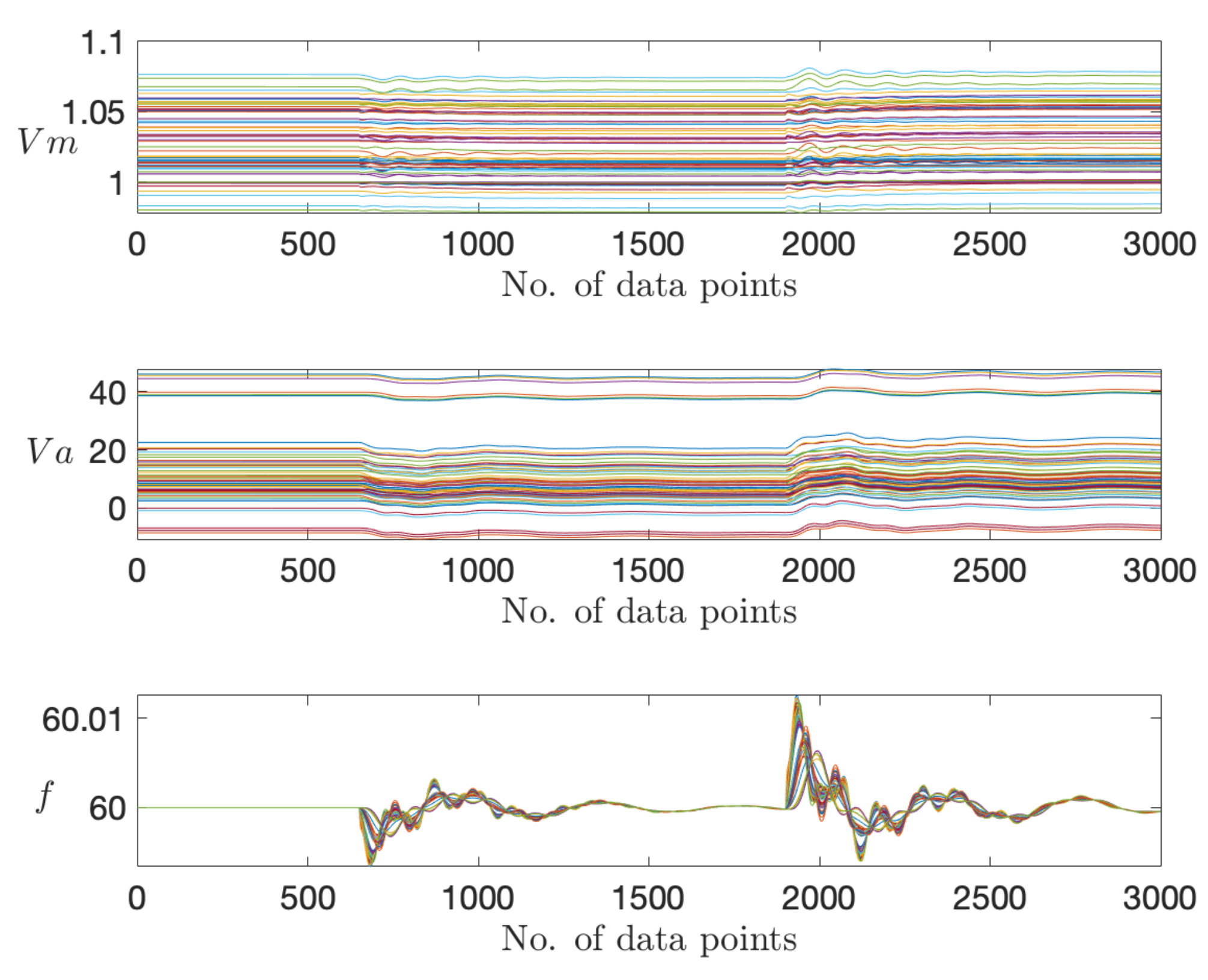}
\caption{Training data using PMU measurements for 3000 time steps.}\label{data_3000}
\end{figure}

\begin{figure}[h!]
\centering
\includegraphics[scale=.4]{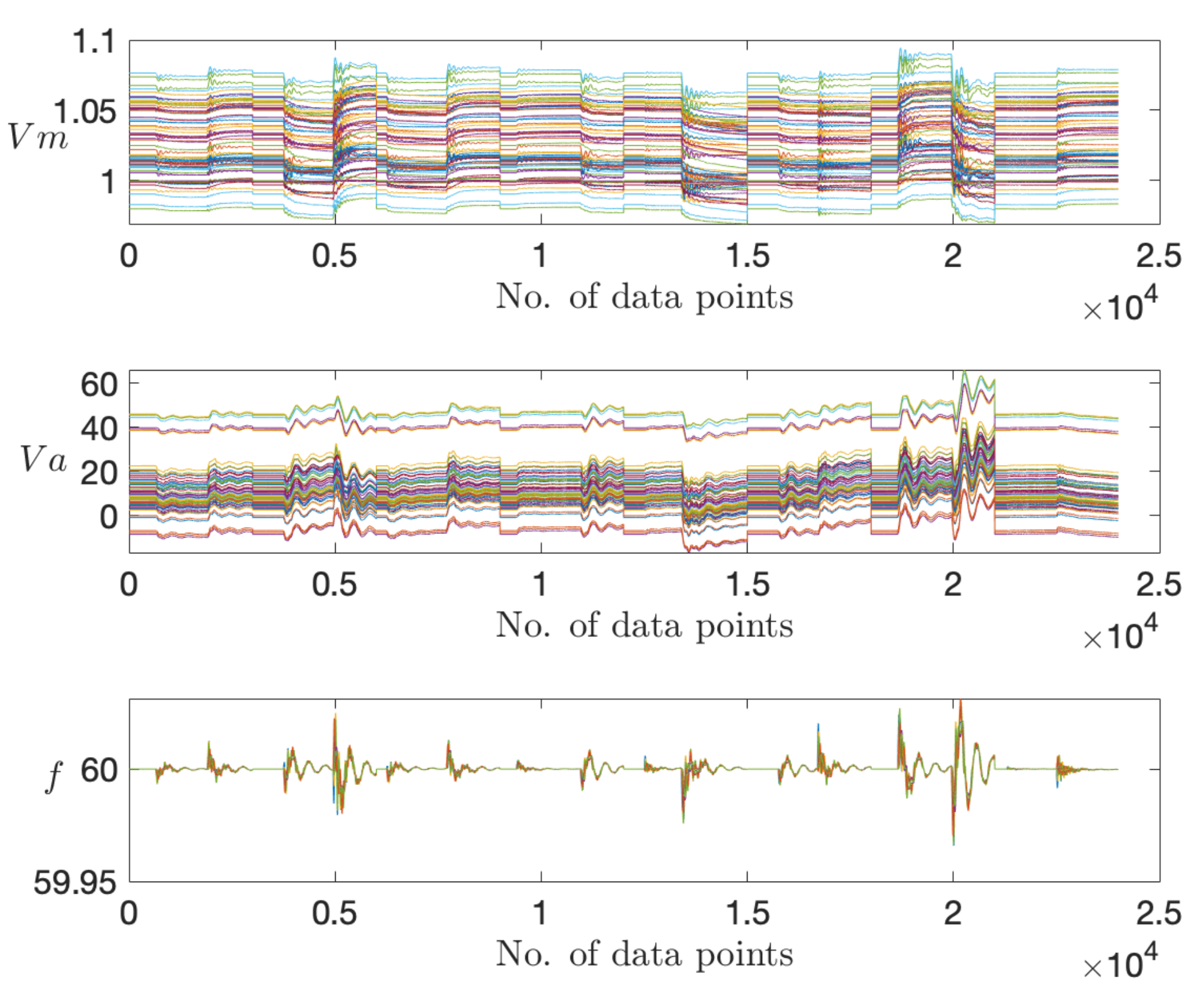}
\caption{Training data using PMU measurements for 24000 time steps.}\label{data_24000}
\end{figure}

The high-level objective here is to identify the evolution of power system dynamics in a purely data-driven fashion applying Koopman operator theory. Specifically, we chose the nodal measurements of PMUs voltage magnitudes $(V_m)$, voltage angles $(V_a)$ and frequencies $(f)$ at each bus to learn an equivalent representation of the power system dynamics in terms of a Koopman operator. Sample time-domain trajectories (data-set used to compute the different Koopman operators) of the variables are shown in Figs. \ref{data_3000} and \ref{data_24000}. 

As discussed in Eq. \eqref{EDMD_formula}, it is seen that a pseudo-inverse needs to be computed to solve for the Koopman operator. As there are $68$ buses and each bus has $3$ states (voltage magnitude, voltage angle and frequency), the combined state space of the system is $204$ $(3\times 68)$. To demonstrate the proposed computation of the Koopman operator in a computationally efficient manner, we consider the PMU data for $8$ seconds of simulation where there are at most $2$ load changes every second. Each second of PMU data contains $3000$ time points and hence the total length of PMU time-series data for $8$ seconds is $24000$. Now, 8 different Koopman operators are computed from 8 different data-sets. The size of the data set is varied from 3000 time steps to 24000 time steps in steps of 3000 time points.  

\begin{figure}
\centering
\includegraphics[scale=.2]{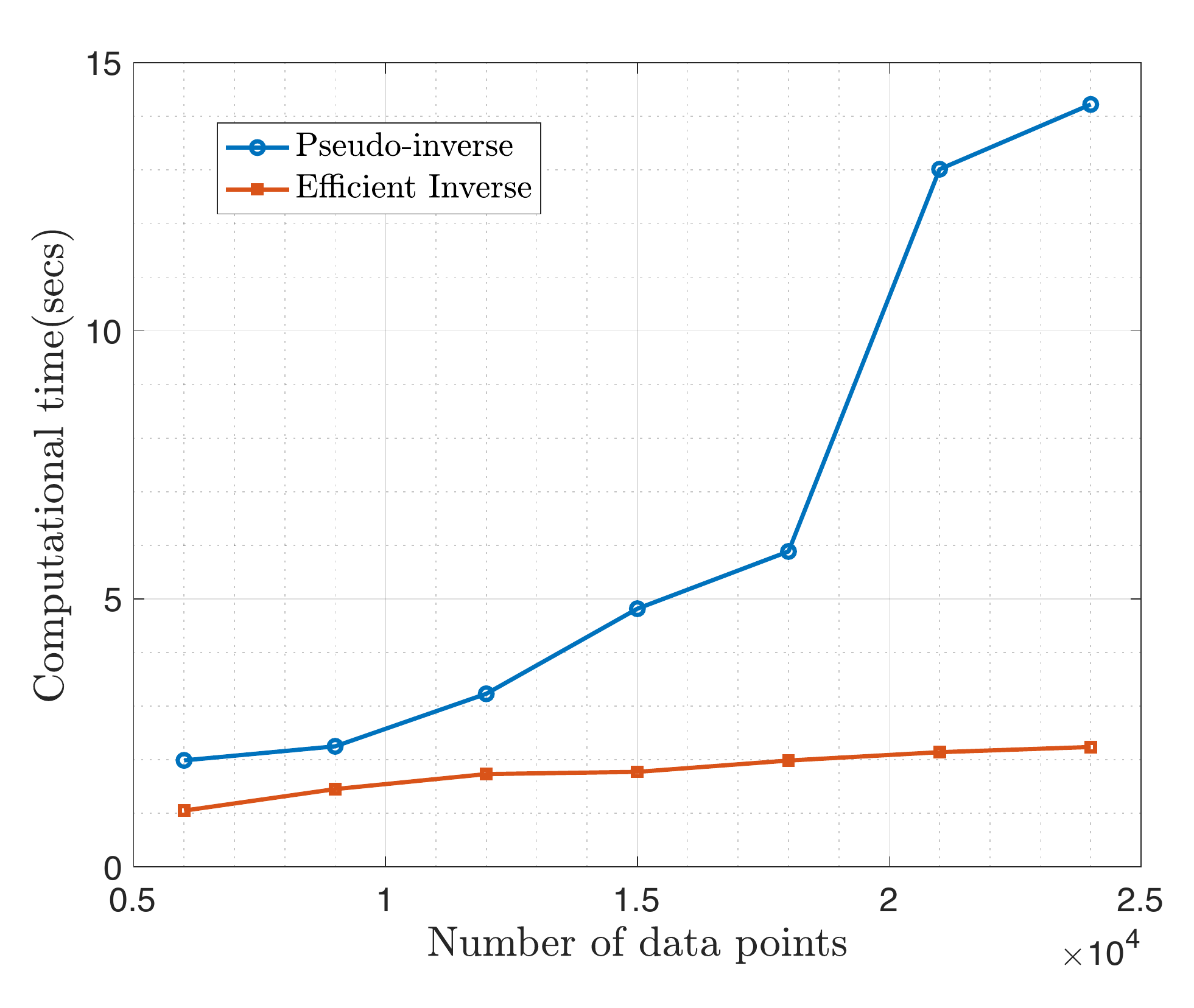}
\caption{Comparison of computation times of the Koopman operator for the power network with different algorithms.}\label{power_comp_time}
\end{figure}

EDMD algorithm with $1000$ Gaussian radial basis functions (RBFs) is considered and hence the corresponding size of the Koopman operator is $1000 \times 1000$. The computation times of the Koopman operators are shown in Fig. \ref{power_comp_time}. As in the case of the oscillator network, it can be seen that the computation time of the Koopman operator using the proposed method is substantially lower than the SVD based ({\tt pinv}) method. Moreover, the trend for the proposed approach is linear and as such, this method will scale much more efficiently compared to the existing method. 

\begin{figure}[h!]
\centering
\subfigure[]{\includegraphics[scale=.225]{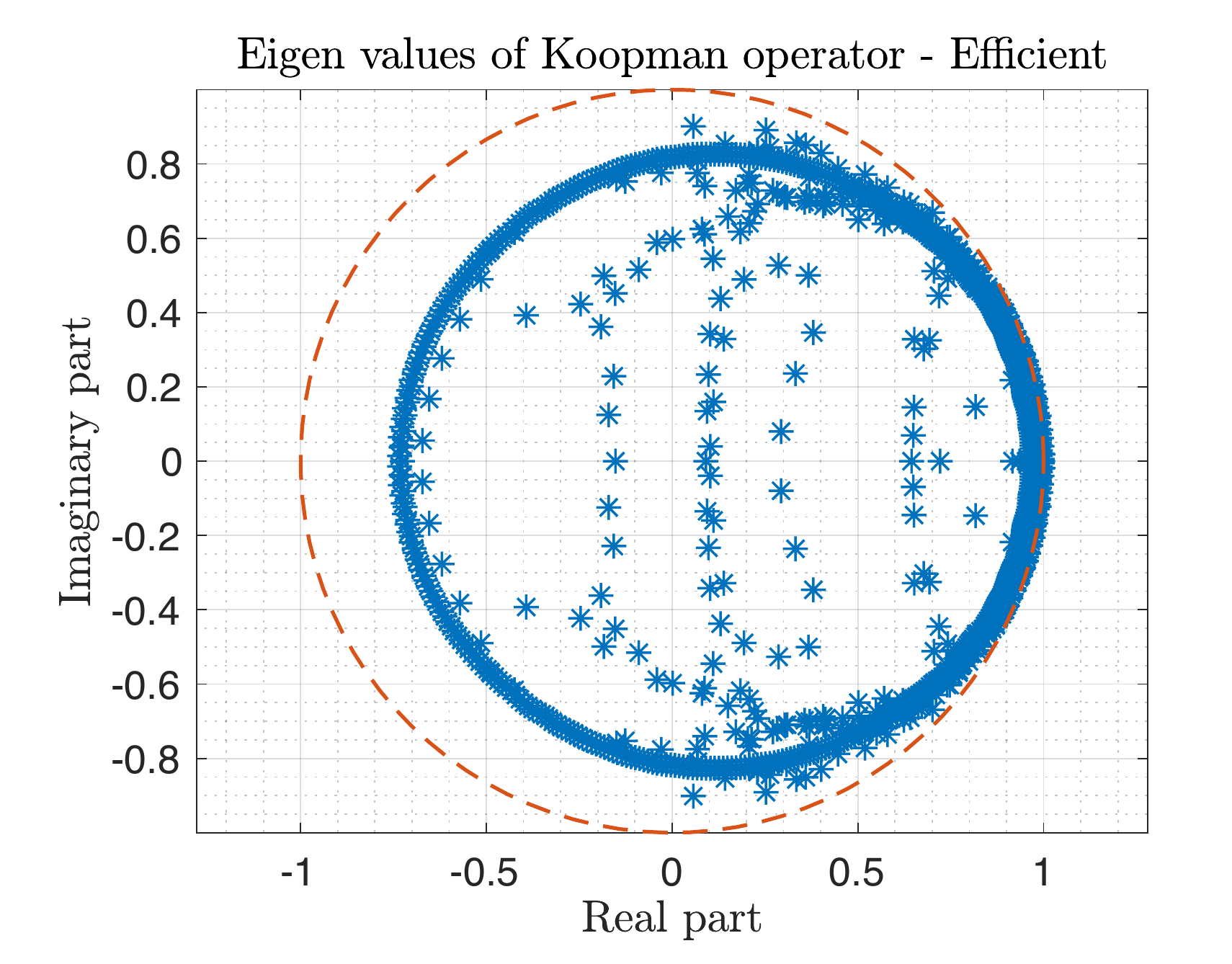}}
\subfigure[]{\includegraphics[scale=.225]{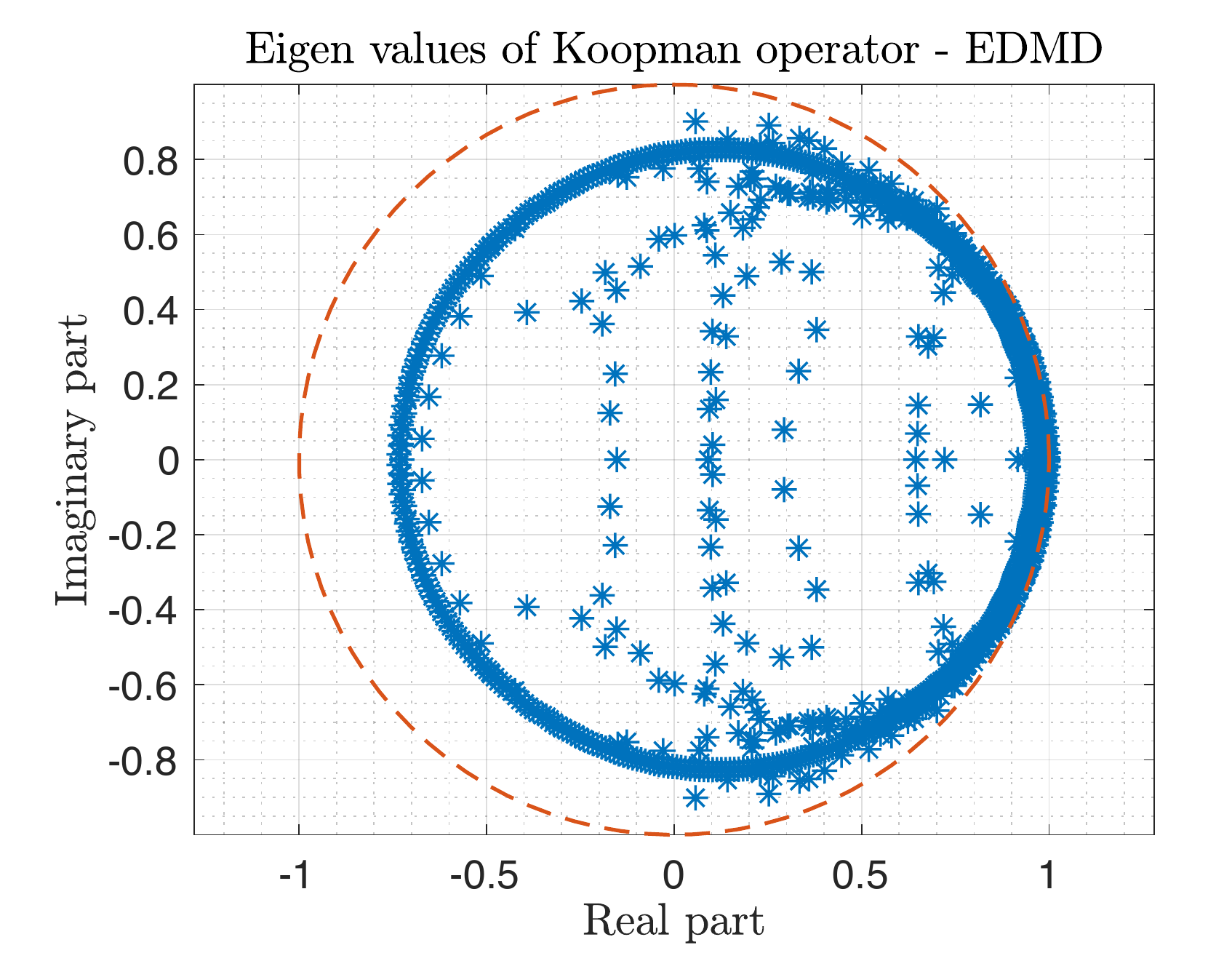}}
\caption{(a) Eigenvalues of the Koopman operator when computed using proposed approach. (b) Eigenvalues of the Koopman operator computed using the SVD based inverse computation ({\tt pinv} in MATLAB).}\label{power_eig}
\end{figure}

Furthermore, in Fig. \ref{power_eig} we plot the eigenvalues of the two Koopman operators obtained by the two different methods and observe that the eigenspectrum is almost similar and thus verifying that the proposed method not only computes the Koopman operator in a time-efficient way, but it does compute the correct Koopman operator.

\section{Conclusions}\label{section_Conclusions}
In this paper, we provide an algorithm for time-efficient identification of large scale dynamical systems from time-series data. In particular, the algorithm efficiently computes the finite-dimensional approximation of the Koopman operator associated with a dynamical system. In general, the computation of the Koopman operator involves the computation of Moore-Penrose inverse of a matrix and if the state-space dimension is large, the computation of the inverse is computationally expensive. In this paper, we discuss a new algorithm based on Cholesky decomposition of the matrix to be inverted, which reduces the computation time significantly. The proposed approach's efficiency is further demonstrated on two different systems, namely, a network of coupled oscillators and the IEEE 68 bus system.

\bibliographystyle{IEEEtran}
\bibliography{ref,ref1,reference}

\end{document}